# On the Evaluation of Neural Code Translation: Taxonomy and Benchmark

Mingsheng Jiao, Tingrui Yu, Xuan Li, Guanjie Qiu, Xiaodong Gu*, Beijun Shen

*School of Software, Shanghai Jiao Tong University, Shanghai, China*
{jiaomingsheng, hzfsls, riken01, qiuguanjie, xiaodong.gu, bjshen}@sjtu.edu.cn

*Abstract*—In recent years, neural code translation has gained increasing attention. While most of the research focuses on improving model architectures and training processes, we notice that the evaluation process and benchmark for code translation models are severely limited: they primarily treat source code as natural languages and provide a holistic accuracy score while disregarding the full spectrum of model capabilities across different translation types and complexity. In this paper, we present a comprehensive investigation of four state-of-the-art models and analyze in-depth the advantages and limitations of three existing benchmarks. Based on the empirical results, we develop a taxonomy that categorizes code translation tasks into four primary types according to their complexity and knowledge dependence: token level (type 1), syntactic level (type 2), library level (type 3), and algorithm level (type 4). We then conduct a thorough analysis of how existing approaches perform across these four categories. Our findings indicate that while state-of-the-art code translation models excel in type-1 and type-2 translations, they struggle with knowledge-dependent ones such as type-3 and type-4. Existing benchmarks are biased towards trivial translations, such as keyword mapping. To overcome these limitations, we construct G-TransEval, a new benchmark by manually curating type-3 and type-4 translation pairs and unit test cases. Results on our new benchmark suggest that G-TransEval can exhibit more comprehensive and finer-grained capability of code translation models and thus provide a more rigorous evaluation. Our studies also provide more insightful findings and suggestions for future research, such as building type-3 and type-4 training data and ensembling multiple pre-training approaches.

## I. Introduction

Code translation aims to port source code from one programming language to another. Software is usually developed on multiple platforms, requiring the same program to be developed in multiple programming languages. Manually translating these programs is tedious and expensive. With the development of deep learning, more and more efforts have been put into neural code translation study to assist programmers with code translation tasks. Pre-training models [1–6] have significantly improved code translation accuracy and become mainstream in the code translation field using a variety of pre-training objectives that focus on different aspects of source code [2, 6, 7] to provide a more comprehensive representation of the source code.

* Xiaodong Gu is the corresponding author

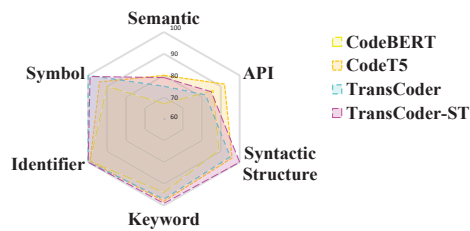

Fig. 1: Fine-grained performance of code translation models in six dimensionalities. We experiment with the Java→C++ translation. The values in the figure are the accuracy obtained on the original dataset.

While current studies focus on improving model architectures and training processes [4, 7, 8], we notice that the evaluation processes and benchmarks for code translation models are deeply limited. First, existing benchmarks such as CodeXGLUE [9] treat all translation tasks to be equally difficult. Second, the current evaluation processes [10, 11] only provide a single score that indicates the overall performance, and such measurements are inevitably tangled with other aspects of model ability such as the translation difficulty. For example, translations that only need to map keywords in different languages can easily achieve high accuracy, while hard translation pairs are heterogeneous in algorithms and can hardly achieve high scores.

In this paper, we perform an in-depth analysis of the fine-grained performance of state-of-the-art models on commonly used benchmarks. We find that state-of-the-art code translation models [1–3, 8] exhibit varying translation capabilities in fine-grained aspects, with a greater proficiency in translating tokens, followed by syntactic structures, libraries, and algorithms. By inspecting translation pairs that achieve different levels of BLEU scores, we observe that different translation pairs have different complexity, requiring different model abilities. Existing benchmarks are biased towards simple translations, such as keyword mapping, while limited in demonstrating model abilities in complex translations, such as libraries and algorithms.

Based on the empirical findings, we develop a taxonomy that categorizes code translation tasks into four types with increasing complexity: **token level (type 1)**, which only needs to learn trivial token-to-token mappings; **syntactic level**

**(type 2)**, which needs to migrate syntactic structures with the knowledge of linguistic rules of both languages; **library level (type 3)**, which depends on external knowledge of libraries and their API usages; and **algorithm level (type 4)**, which needs to rewrite the input code with different algorithms. Our further experiments confirm that while existing approaches achieve superb performance on type-1 and type-2 translations, they struggle with knowledge-dependent ones such as type-3 and type-4.

Based on the taxonomy and empirical results, we propose a new benchmark called G-TransEval by carefully curating type-3 and type-4 translations and unit tests. Our experiments on the new benchmark indicate that G-TransEval can exhibit a more comprehensive and unbiased capability of code translation models and thus provide a more rigorous evaluation. The results have also yielded several insightful findings, such as (1) unsupervised models perform better on low-level translations, whereas supervised models achieve better results on high-level translations; (2) translation difficulty varies across different programming languages, with dynamic-to-static translations being the most challenging; (3) higher-level translations rely more heavily on large-scale training data; and (4) LLMs alleviate the knowledge gap of type-2 and type-3 translations through ultra-large training data and model parameters. Our work also provides new insights that could help advance and evaluate code migration systems in the future. For instance, building type-3 and type-4 training data, and ensembling multiple pre-training approaches.

Our contributions can be summarized as follows:
- We empirically investigate the fine-grained ability of state-of-the-art code translation models and analyze the weakness of existing benchmarks.
- We are the first to develop a taxonomy that categorizes code translation tasks into four types according to their complexity and knowledge dependence. The taxonomy can successfully capture the complexity of various code translation types.
- We propose a new benchmark by curating high-quality test samples across different languages and translation types. Results show that our benchmark can exhibit more comprehensive and finer-grained capabilities of current code translation models, providing new insights on future work.

## II. EMPIRICAL STUDY

### A. Research Questions

**RQ1. How is the fine-grained performance of state-of-the-art code translation models?**

We perform an in-depth analysis of the fine-grained performance of SOTA code translation models. Specifically, we manually examine the translation accuracy (i.e., the percentage of correct translations) in terms of six dimensionalities:

1) **Keywords**: keyword mappings between programming languages such as "boolean→bool" in Java→C++ translation.
2) **Identifiers**: variable and function names. As renaming does not affect the semantics, we allow for variable renaming in the translation and only consider misused identifier names (e.g. should be $j$ but got $i$ instead) as incorrect translations.
3) **Symbols**: syntactic symbols in source code such as punctuation, brackets, and operators.
4) **Syntactic Structures**: sequences of syntactically related tokens [12]. We consider a translation correct if it does not violate syntactic rules.
5) **APIs**: API mappings for libraries. Translations using APIs different from golden answers while with the same meaning, such as `str.length()` and `str.size()`, `map.find(e) != map.end()` and `map.count(e)>0`, are also marked as being correct.
6) **Semantic**: the functionality of the code. We measure the accuracy using the passing rate on the unit tests.

**RQ2. Can existing benchmarks exhibit the fine-grained capability of code translation models?**

We perform code translation on three popular benchmarks and partition the test code pairs into groups based on their BLEU levels. For each group, we manually examine the characteristics of the translation code. This allows us to scrutinize the specific attributes of translation pairs that yield varying performance levels and gain a deeper understanding of how existing models fare under different degrees of translation complexity.

### B. Models

We evaluate the performance of four pre-trained code translation models that involve both supervised and unsupervised methods [4].

**CodeBERT** [1] is a widely-used pre-trained model for source code, which is built on the Transformer encoder. The model is pre-trained on two self-supervised objectives, namely, MLM (masked language model) and RTD (replaced token detection), with bimodal data in both natural and programming languages. As an encoder-only model, an additional decoder is required to be trained when we apply it to code translation.

**CodeT5** [3] is a pre-trained encoder-decoder model for source code based on T5 [13]. The model is pre-trained with identifier-aware objectives to capture token-type information in programming languages. Besides, a bimodal dual generation objective is utilized for bidirectional NL-PL conversions to improve the generation ability of the decoder.

**TransCoder** [2] is an unsupervised code translation model that is pre-trained on a large amount of monolingual source code. TransCoder pre-trains both the encoder and decoder using denoising auto-encoding (DAE) and back translation (BT) objectives. DAE aims to train the model's generation capacity, while BT aims to produce high-quality translation results.

**TransCoder-ST** [8] is an unsupervised code translation model that extends TransCoder with an automated unit test generation pipeline. The extension reduces noise in the back-translation process and filters out invalid translated code. TransCoder-ST substantially improves TransCoder and DOBF [5].

### C. Datasets

**TransCoder-test** [2] refers to the test set of TransCoder, which was originally proposed for evaluating TransCoder. The dataset involves 1,418 parallel samples across C++, Java, and Python collected from multilingual problem solutions in GeeksForGeeks[1]. The samples are split into valid and test sets with 470 and 948 samples, respectively. TransCoder-test is a desirable test set for the reason that half of the translation pairs are equipped with unit tests.

**CodeXGLUE** [9] is a widely used benchmark for training and evaluating code intelligence tasks. CodeXGLUE contains 14 datasets, including a code translation dataset between Java and C#, with 10,253/499/1000 examples for training, validation, and testing, respectively. The parallel code is collected from several bilingual projects, including Lucene[2], POI[3], JGit[4], and Antlr[5].

**XLCoST** [14] is a large-scale corpus extracted from multilingual problem solutions in GeeksForGeeks. The benchmark involves fine-grained parallel code (1 million parallel snippets and 123K parallel programs) across seven programming languages (i.e., C++, Java, Python, C#, JavaScript, PHP, and C), and natural language (English). In this paper, we focus on method-level translations. Thus, we extract parallel functions from the program-level parallel code. The final dataset contains 9,889 parallel tuples of functions. We split them into training, validation, and test sets with 8,389/500/1000 samples, respectively.

### D. Evaluation Metrics

We evaluate code translation models using three of the most popular metrics:

**BLUE** [10] is a commonly used metric for evaluating the quality of machine translations. It measures the ratio of n-grams overlap between the translation and the reference [10]. We use the BLEU-4 score which calculates the weighted average of BLEU scores with 1-4 grams. During our experiments, we encounter an issue where a sequence of tokens may stick into a single token, causing incorrect BLEU scores. For instance, `Array.sort()` would be treated as a single token in BLEU calculation, even though it contains five tokens. To address this issue, we preprocess all code by inserting spaces between all tokens. For instance, "`Array.sort()`" is stretched into "`Array . sort ( )`".

**CodeBLEU** [11] is a metric that is specifically designed for evaluating code generation tasks. In addition to weighted n-gram matching, CodeBLEU ensembles two more matchings of syntactic trees and semantic data flows, which give a more comprehensive assessment of the translation quality than solely n-gram matching.

**CA (Computational Accuracy)** [2] is a metric to measure the functional correctness of machine-generated code. Unlike text-based metrics such as BLEU, CA matches the functionality-level equivalence of code using the pass rate of unit tests. Hence it is more reliable in distinguishing the correctness of generated code.

### E. Implementation Details

For all evaluated models, we download their official checkpoints[6,7,8]. In the subsequent experiments, the models are configured as follows: TransCoder and TransCoder-ST are used directly with the downloaded checkpoints, while CodeT5 and CodeBERT are fine-tuned on the training set of XLCoST, except when evaluating on the CodeXGLUE dataset, where the models are fine-tuned on CodeXGLUE's own training set. Each model is fine-tuned for 10 epochs, and the checkpoint that achieves the highest BLEU score on the validation set is used for evaluation.

All hyperparameters, as summarized in Table I, are consistent with the open-source documents provided in the original paper of each model. We train and evaluate all models on 2 × NVIDIA GeForce RTX 4090.

TABLE I: Hyperparameters of the evaluated models.

| Hyperparameter | CodeBERT | CodeT5 | Transcoder(-ST) |
|---|---|---|---|
| # of Transformer layers | 12 | 24 | 12 |
| Max length of position | 512 | 512 | 512 |
| Embedding size | 768 | 768 | 1024 |
| Attention head | 12 | 12 | 8 |
| Vocabulary size | 50,265 | 32,100 | 64,001 |
| # of parameters | 125M | 220M | ∼ 100M |
| Train batch size* | 8 | 8 | / |
| Learning rate* | 5e-5 | 5e-5 | / |
| Max input length | 512 | 512 | 512 |
| Max output length | 512 | 512 | 512 |
| Beam size | 5 | 5 | 5 |

\* Here, train batch size and learning rate are used during fine-tuning.

## III. RESULTS AND ANALYSIS

### A. RQ1: Fine-grained Performance of SOTA Code Translation Models

We evaluate the accuracy of Java→C++ translation for SOTA code translation models on the TransCoder-test benchmark. We select 100 samples with unit tests from the test set and use them to examine the fine-grained performance of CodeBERT, CodeT5, TransCoder, and TransCoder-ST. Three authors independently examine the translations generated by each model and mark the correctness according to the definition of each dimensionality. Conflicts are settled down via majority voting.

The results are presented in Figure 1. First, all models demonstrate exceptional performance in translating keywords and identifiers, achieving an average accuracy of 98.5%. Subsequently, all models exhibit high accuracy in translating syntactic structures and symbols, achieving over 97% accuracy on both dimensions. It is worth noting that CodeBERT's accuracy is slightly lower, at 89% and 90%, for syntactic structures and symbols respectively, which may be attributed to the absence of a pre-trained decoder. Furthermore, generating accurate APIs is comparatively more challenging, with an average accuracy score of 86.25%. Lastly, semantic translations (i.e., unit tests pass rate) have been proven to be the most challenging task, which achieves an average accuracy of only 75.25%. The results indicate that different code elements possess varying levels of translation complexity, and that current code translation models exhibit different levels of proficiency in translating them.

---

[1] https://www.geeksforgeeks.org/
[2] http://lucene.apache.org/
[3] http://poi.apache.org/
[4] https://github.com/eclipse/jgit/
[5] https://github.com/antlr/

[6] https://huggingface.co/microsoft/codebert-base
[7] https://huggingface.co/Salesforce/codet5-base
[8] https://github.com/facebookresearch/CodeGen

**Answer to RQ1:** State-of-the-art code translation models exhibit varying translation capabilities in fine-grained aspects, with a greater proficiency in translating tokens, followed by syntax, APIs, and semantics.

### B. RQ2: Distinguishing Ability of Existing Benchmarks

Having noticed the fine-grained performance of various models, we analyze whether existing benchmarks can distinguish the fine-grained performance. Specifically, we fine-tune a CodeT5 model using the experimental setup in Section II-E and then test the performance using the three benchmarks mentioned in Section II-C. We examine Java→C# translations for CodeXGLUE and Java→C++ translations for both TransCoder-test and XLCoST.

We start by examining the translation results on the CodeXGLUE benchmark. To our surprise, we find that out of the 1,000 translation results, 681 exhibited exact matches with the reference translations. Notably, we observe a frequently occurrent pattern handling HTTP requests, where translation pairs share completely identical processing, except for the prefix of request types. This pattern constitutes a large portion of over 1/4 in both the training and test sets, causing the model to easily achieve 100% accuracy on this translation pattern. These duplicated pairs have a negative impact on the quality of the benchmark and consequently affect the evaluation of the model's translation ability.

We observe a similar trend on TransCoder-test and XLCoST. For example, XLCoST has 1,000 test samples, where 511 can be exactly matched by CodeT5's translation. CodeT5 also produces 305 exactly-matched translations out of the 948 test examples in TransCoder-test. This indicates that the distinguishing ability of benchmarks is threatened by the quality of test examples.

To further analyze the distinguishing ability of benchmarks, we group all the generated results according to their BLEU scores (>80, 50-80 and <50) and explore the characteristics of translation pairs within each group. The results are summarized in Table II. Across all three benchmarks, a significant proportion of translations in the test set achieves high BLEU

TABLE II: Characteristics of code under different groups of BLEU scores. The three colors (yellow, green, and blue) stand for TransCoder-test, XLCoST, and CodeXGLUE, respectively.

| BLEU | Percentage | Characteristics of Code |
|---|---|---|
| 80-100 | 84.4 / 88.7 / 83.4 | Basic data types<br>Simple condition statements<br>Arithmetic operations<br>Simple function calls |
| 50-80 | 13.4 / 10.9 / 12.0 | Basic data structures<br>Diverse conditions statements<br>Fewer arithmetic operators<br>Multiple API calls |
| 0-50 | 2.2 / 0.4 / 4.6 | Complex variable types<br>Longer and informative identifiers<br>Manipulation of complex variables<br>Complex API calls<br>Difference in algorithm |

scores, whereas the proportion of translations with low BLEU scores is small.

Upon manual examination, we find that translations with higher BLEU scores tend to be short and trivial functions: they use only basic data types and short and uninformative identifiers. The functions' functionality is usually expressed through straightforward arithmetic operations that are consistent across different languages. Simple data structures are used without complex branches and loops. These functions constitute the majority of the test set and can achieve a BLEU score of 100 or near it, contributing greatly to the overall model score. As the length of functions increases, more complex data structures such as vectors, maps, and sets appear in the functions, leading to an evident decrease in BLEU scores. This is probably because longer identifiers and more complex data structures carry more diverse semantics, increasing the heterogeneity between source and target languages. The invoke of non built-in libraries (APIs with parameters) further increases the difficulty of translation.

In summary, translations that achieve high BLEU scores tend to be simple and trivial function pairs, which use only basic variable types and involve simple variable operations. As the complexity of the function increases, such as the introduction of complex data structures and library APIs, the BLEU score tends to decrease. Differences in algorithm can also lead to a decrease in BLEU score as they introduce complex logic mappings, thereby increasing the difficulty of translation.

**Answer to RQ2:** Existing benchmarks are biased towards trivial translations, such as token mapping, and are limited in complex translations, such as library invocation and algorithm rewriting.

## IV. TAXONOMY

The empirical results reveal that the state-of-the-art benchmarks have a bias towards trivial translations, which is a significant reason for the decrease of the discriminative ability

TABLE III: Taxonomy of code translation tasks. *Internal* means requiring language-specific knowledge only, while *External* means requiring additional knowledge beyond language syntax. The symbol ○ means optional.

| Taxonomy | Description | Definition | Knowledge Dependence | | | |
| --- | --- | --- | --- | --- | --- | --- |
| | | | Internal | | External | |
| | | | Keyword | Grammar | API | Algorithm |
| Type 1 | Token-level translation | Map trivial tokens to their equivalent in the target | ✓ | ✗ | ✗ | ✗ |
| Type 2 | Syntax-level translation | Migrate syntactic structures based on linguistic rules | ○ | ✓ | ✗ | ✗ |
| Type 3 | Library-level translation | Migrate library to their equivalent in the target language | ○ | ○ | ✓ | ✗ |
| Type 4 | Algorithm-level translation | Reimplement the program in the target language using a different algorithm | ○ | ○ | ○ | ✓ |

TABLE IV: Examples for the four translation types. The translation for each type is highlighted in yellow.

| | C++ | Java |
| --- | --- | --- |
| TYPE 1 | ```cpp
int maxProductSubset(int a[], int n){
  if(n == 1) return a[0];
  int max_neg = INT_MIN;
  int prod = 1;
  for(int i = 0; i < n; i++) {
    //... Rest of the Code
  }
  return prod;
}
``` | ```java
int maxProductSubset(int a[], int n) {
  if(n == 1) return a[0];
  int max_neg = Integer.MIN_VALUE;
  int prod = 1;
  for(int i = 0; i < n; i++)
    //... Rest of the Code
  }
  return prod;
}
``` |
| TYPE 2 | ```cpp
int countWays(string s) {
  int count[26] = {0};
  for(char x : s)
    count[x - 'a']++;
  count[s[0] - 'a'] = 1;
  int ans = 1;
  //... Rest of the Code
  return ans ;
}
``` | ```java
int countWays(String s) {
  int count[] = new int[26];
  for(int i = 0; i < s.length(); i++)
    count[s.charAt(i)-'a']++;
  count[s.charAt(0)-'a']=1;
  int ans = 1;
  //... Rest of the Code
  return ans ;
}
``` |
| TYPE 3 | ```cpp
int removeConsecutiveSame(vector<string> v){
  stack<string> st;
  for(int i = 0; i < v.size(); i++){
    if(st.empty()) st.push(v[i]);
    else {
      string str = st.top();
      if(str.compare(v[i])==0) st.pop();
      else st.push(v[i]);
    }
  }
  return st.size();
}
``` | ```java
int removeConsecutiveSame(Vector<String> v){
  Stack<String> st = new Stack<>();
  for(int i = 0; i < v.size(); i++){
    if(st.empty()) st.push(v.get(i));
    else {
      String str = st.peek();
      if(str.equals(v.get(i))) st.pop();
      else st.push(v.get(i));
    }
  }
  return st.size();
}
``` |
| TYPE 4 | ```cpp
int calFactorial (int n){
  int result = 1;
  for(int i=1; i<=n; ++i)
    result *= i;
  return result;
}
``` | ```java
int calFactorial (int n){
  if(n == 1 || n == 0) return 1;
  return n*calFactorial(n-1);
}
``` |

by BLEU scores [2, 11, 15–18]. In this section, we develop a taxonomy of code translation tasks regarding translation complexity to provide a more fine-grained evaluation of model performance. Specifically, we categorize code translation into four levels regarding their translation complexity and knowledge dependence:

**Type 1: Token level translation.** The source and target code differ only in trivial tokens such as keywords and identifiers. Type-1 translation is straightforward since the target code can be obtained by minimal token deletion, insertion, and replacement, without requiring linguistic rules of the programming language.

**Type 2: Syntax level translation.** The source and target code differ in syntactic structures, e.g., the way of variable declaration, list/string iteration, and type casting. Type-2 translations are more challenging than type-1 because it relies on the linguistic rules of a programming language and the context of the code.

**Type 3: Library level translation.** The source and target code differ in the usage of libraries. In addition to language-specific knowledge, type-3 translation also needs external knowledge about the libraries, for instance, what is the functionality of an API and what the arguments mean. Hence, it is considered to be more challenging than type-2 translations.

**Type 4: Algorithm level translation.** Despite equivalent functionalities, the source and target code is largely disparate in algorithms and logic. The token repetition rate between the two is minimal, such as different algorithms implemented in two languages with unique features (e.g., the collection iterator in C++ *vs.* the Stream class in Java) or implementing a specific API in the source language that has no counterpart in the target language (e.g., "`zip()`" and "`enumerate()`" in Python).

Table III presents a comparison of the four types of code translation tasks based on their definitions and dependence on various types of knowledge. It is important to note that if multiple knowledge sources are necessary for a particular translation, the type is determined based on the most challenging level of knowledge dependence. For example, a translation sample implemented using different algorithm can still be categorized as type-4 even if it does not meet the requirements of type-3.

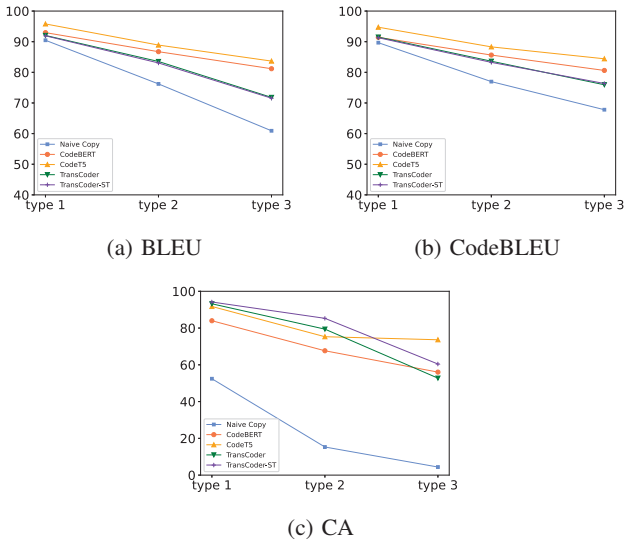

(a) BLEU  (b) CodeBLEU  (c) CA

Fig. 2: Performance of various models under different translation types. We evaluate Java→C++ translations on the categorized TransCoder-test benchmark.

Table IV provides examples of the four types of code translation between C++ and Java.

To validate the efficacy of our taxonomy, we categorize the TransCoder-test data according to our taxonomy and conduct a series of experiments based on it. Due to the absence of type-4 translations in TransCoder-test, we divide its data into three categories, consisting of 384/341/223 samples, corresponding to type-1, 2, and 3 respectively.

Figure 2 presents the performance of various models under different translation types. It can be observed that as the translation level goes up, the performance decreases in all metrics. For example, compared to type-1 translations, TransCoder achieves 9.2% lower score in BLEU and 14.8% lower score in CA on type-2, and 21.8% lower BLEU and 42.5% lower CA on type-3. The same trend can be discovered in other approaches. More surprisingly, we find that even Naive Copy (i.e., the naive approach that simply copies the input code) can achieve BLEU scores of 90 in type-1 translations.

The results suggest that the four categories in our taxonomy indeed differentiate the increasing complexity of code translation, which affirms the validity of our 4-level categorization of code translation tasks.

## V. BENCHMARK

Based on the taxonomy and empirical findings, we propose G-TransEval, a new benchmark for evaluating code translation models. Our benchmark leverages existing benchmarks and extends them to involve all four translation types. To increase the reliability of performance measures, we also augment each sample with unit test cases. The construction involves the following steps.

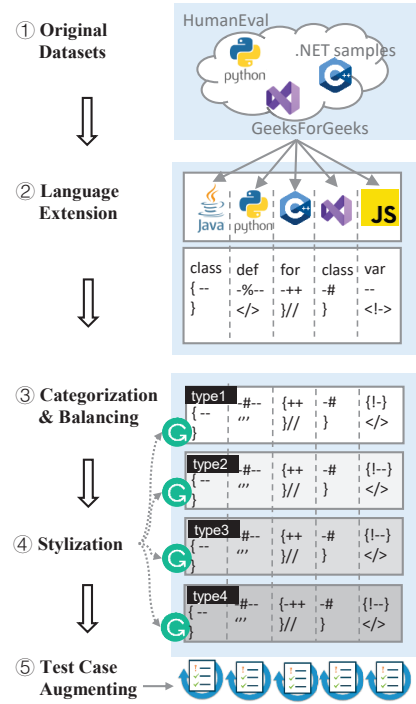

Fig. 3: The construction workflow of G-TransEval

*1) Collecting parallel code from diverse sources:* Previous benchmarks such as XLCoST and TransCoder-test only involve programming problems from GeeksForGeeks. To increase the diversity of translations, we gather parallel code from multiple sources. Except for collecting C++ answers to programming questions from GeeksForGeeks, we also incorporate the HumanEval [19] benchmark, which consists of 164 programming problems written by humans with their Python solutions. Due to the limited availability of code for type-4 translations, we further fetch 25 C# samples from .NET code samples[9].

*2) Language extension:* The initial dataset has only monolingual solutions to each problem. To enable the evaluation of multilingual translations, we manually expand it to five programming languages, i.e., Python, C++, Java, C#, and JavaScript. For each solution to a programming problem, two of the authors independently write equivalent code in the other languages, and resolve any discrepancy through discussion until a consensus is reached.

*3) Categorization and balancing:* One of the main discoveries from our empirical study is that the performance of translation models is significantly affected by translation types within our taxonomy. Consequently, we partition our dataset

[9]https://learn.microsoft.com/en-us/samples/dotnet/try-samples/101-linq-samples/

TABLE V: Comparison between G-TransEval and existing code translation benchmarks (G4G = GeeksForGeeks).

| Dataset | Source | Parallel Data Size (train/valid/test) | Languages | Categorized? | Style Normalized? | Unit Tests Included? | Golden Answer Verified? |
|---|---|---|---|---|---|---|---|
| CodeXGLUE | Lucune, POI, JGit, Antlr | 10,253 / 499 / 1,000 | Java, C# | ✗ | ✗ | ✗ | ✗ |
| XLCoST* | G4G | 9450 / 490 / 901 | C++, Java, C#, PHP, JavaScript, Python, C | ✗ | ✗ | ✗ | ✗ |
| TransCoder-test | G4G | - / 470 / 948 | C++, Java, Python | ✗ | ✗ | Partial | ✗ |
| HumanEval-X | HumanEval | - / - / 164 | C++, Java, Go, JavaScript, Python | ✗ | ✓ | ✓ | ✓ |
| G-TransEval | HumanEval, G4G, .NET samples | - / - / 400 | C++, Java, C#, JavaScript, Python | ✓ | ✓ | ✓ | ✓ |

* XLCoST provides parallel code at both snippet and program levels and has varying numbers of parallel code across different languages. To be consistent with other benchmarks, we show the largest size of the program level parallel language pairs in their GitHub repository.

into four subsets based on the corresponding taxonomy rules. To ensure that each type is adequately represented, we have chosen 125 samples for type-1, 2, and 3 translations, and 25 samples for type-4 translations, while discarding the remaining samples. As a result, our benchmark comprises a total of 400 samples.

*4) Style normalization:* To mitigate the bias towards certain coding styles, we rephrase the code samples following the naming conventions outlined in the Google style guide[10]. We perform the majority of the stylization work manually and also use C++, Java, and JavaScript solutions from HumanEval-X as references [20].

*5) Augmenting unit test cases:* Finally, we equip each sample with unit test cases to increase the reliability of the performance evaluation. Two of the authors independently write the test cases, with their results discussed and checked by all the co-authors.

The golden translations are also manually verified to guarantee that they can pass the unit tests. This step ensures that all parallel code has the same semantics and functionality.

### A. Comparison with Prior Benchmarks

Our benchmark, G-TransEval, is the first categorized test set designed to provide fine-grained and extensive evaluations of code translation models. To showcase the efficacy of G-TransEval, we compare it against other commonly used benchmarks. Table V provides a comprehensive overview of the comparison between various benchmarks in terms of size, languages, inclusion of unit tests, golden answer verification, and categorization. Notably, G-TransEval comprises a total of 400 code translation pairs, with the first three types consisting of 125 translation pairs each, and the fourth type comprising 25 translation pairs.

*1) G-TransEval vs. CodeXGLUE:* The CodeXGLUE involves parallel (Java-C#) code collected from bilingual projects. In order to perform translation on this dataset, project-specific information is required such as the corresponding C# class name of a specific Java class defined by project developers. Additionally, as demonstrated in Section III-B, some special patterns must be recognized by the model to translate correctly between two languages. To obtain this information, it must be learned from a parallel training set, which is not feasible for unsupervised approaches since the training set is not large enough to train a model in unsupervised way. In contrast, G-TransEval comprises multilingual solutions to programming problems that primarily rely on general knowledge of programming languages. Consequently, it is suitable for evaluating both supervised and unsupervised methods.

[10]https://google.github.io/styleguide/

*2) G-TransEval vs. XLCoST:* The XLCoST benchmark comprises parallel programs that solve programming problems from GeeksForGeeks. While it contains a large number of samples, they are unevenly distributed over different translation types. A considerable portion of the samples only involve basic arithmetic operations or straightforward conditions and loops, resulting in relatively high BLEU scores. Comparatively, G-TransEval categorizes samples into four levels based on the complexity and also includes unit tests for all samples, enabling fine-grained and reliable evaluations of model performance. Additionally, the code convention is normalized according to Google style guides, ensuring the fairness of model comparison.

*3) G-TransEval vs. TransCoder-test:* TransCoder-test comprises parallel methods sourced from GeeksForGeeks with unit tests. However, the availability of unit tests is limited in TransCoder-test, as only a subset of its samples are equipped with unit tests, and some translation pairs have unit tests in only one language. Additionally, the golden translations in TransCoder-test have not been verified by unit tests. By contrast, G-TransEval contains unit tests for all samples, and all multilingual golden answers in G-TransEval are verified to pass those tests.

*4) G-TransEval vs. HumanEval-X:* HumanEval-X is an extended version of HumanEval, which contains 164 hand-crafted programming problems, and corresponding solutions in Python, C++, Java, JavaScript, and Go. Each problem is equipped with unit tests. However, HumanEval-X is not categorized and the 164 samples are unevenly distributed over different translation types. Particularly, most of them are type-4 translations (i.e., totally disparate in algorithms). By comparison, G-TransEval involves a balanced distribution of type-1, 2, and 3 samples, which enables a more comprehensive evaluation of translation models. G-TransEval also advances HumanEval-X in that it incorporates samples collected from multiple sources, providing a more diverse and challenging evaluation for translation models.

TABLE VI: Comparison of model performance on different translation types on the proposed benchmark (CB = CodeBLEU).

| Model | Java→Python | | | Python→Java | | | Java→C++ | | | C++→Java | | | Java→JavaScript | | | JavaScript→Java | | |
|---|---|---|---|---|---|---|---|---|---|---|---|---|---|---|---|---|---|---|
| | BLEU | CB | CA | BLEU | CB | CA | BLEU | CB | CA | BLEU | CB | CA | BLEU | CB | CA | BLEU | CB | CA |
| **Type 1** | | | | | | | | | | | | | | | | | | |
| CodeBERT | 81.37 | 82.67 | 78.40 | 81.26 | 84.14 | 67.20 | 93.83 | 94.18 | 83.20 | 95.68 | 95.54 | 84.00 | 83.77 | 84.91 | 78.40 | 93.47 | 93.52 | 72.80 |
| CodeT5 | 82.71 | 83.07 | 88.00 | 81.98 | 84.81 | 78.40 | 94.14 | 94.41 | 90.40 | 97.39 | 97.35 | 94.40 | 84.67 | 85.25 | 85.60 | 93.46 | 93.81 | 76.80 |
| TransCoder | 86.28 | 83.73 | 57.60 | 82.82 | 85.32 | 78.40 | 89.73 | 90.61 | 94.40 | 93.55 | 94.22 | 92.80 | - | - | - | - | - | - |
| TransCoder-ST | **90.12** | **88.66** | 80.80 | **90.86** | **91.94** | **88.00** | 87.95 | 88.78 | **97.60** | 94.50 | 95.15 | **95.20** | - | - | - | - | - | - |
| **Type 2** | | | | | | | | | | | | | | | | | | |
| CodeBERT | 77.52 | 77.70 | 53.60 | **69.75** | 68.04 | 33.60 | 89.19 | 89.17 | 62.40 | 77.83 | 76.30 | 55.20 | 80.22 | 79.57 | 55.20 | 77.02 | 74.01 | 42.40 |
| CodeT5 | 78.90 | 79.11 | 74.40 | 69.57 | **68.91** | 50.40 | **91.33** | **91.47** | **72.80** | **82.15** | **81.28** | **83.20** | **82.74** | **82.06** | 73.60 | **80.35** | **78.07** | **62.40** |
| TransCoder | 84.39 | 84.37 | 60.80 | 65.13 | 65.18 | 46.40 | 83.76 | 84.95 | 69.60 | 69.50 | 68.77 | 57.60 | - | - | - | - | - | - |
| TransCoder-ST | **87.38** | **87.60** | **76.00** | 66.11 | 64.82 | **51.20** | 83.93 | 84.85 | 71.20 | 70.11 | 69.55 | 55.20 | - | - | - | - | - | - |
| **Type 3** | | | | | | | | | | | | | | | | | | |
| CodeBERT | 74.51 | 74.38 | 28.80 | 63.69 | 62.96 | 16.80 | 79.14 | 80.64 | 31.20 | 71.81 | 69.49 | 26.40 | 72.56 | 72.13 | 25.60 | 72.03 | 68.56 | 20.00 |
| CodeT5 | 78.62 | 78.95 | 68.00 | 67.47 | 67.61 | **44.80** | **83.66** | **84.67** | **48.00** | **74.52** | **74.26** | **58.40** | 75.18 | 75.71 | **52.80** | 74.26 | 72.04 | **37.60** |
| TransCoder | 78.04 | 77.71 | 26.40 | 65.00 | 63.76 | 19.20 | 74.83 | 78.02 | 38.40 | 68.86 | 66.27 | 33.60 | - | - | - | - | - | - |
| TransCoder-ST | **84.42** | **84.15** | **69.60** | **70.84** | **69.14** | 38.40 | 76.20 | 79.26 | 40.80 | 68.45 | 67.69 | 36.80 | - | - | - | - | - | - |
| **Type 4** | | | | | | | | | | | | | | | | | | |
| CodeBERT | 35.89 | 37.10 | 0.00 | 26.32 | 30.22 | 0.00 | 25.18 | 33.53 | 0.00 | 20.38 | 28.05 | 0.00 | 34.05 | 36.48 | 0.00 | 22.88 | 27.93 | 0.00 |
| CodeT5 | 37.08 | 39.82 | 0.00 | 21.79 | 28.13 | 0.00 | **31.76** | **40.87** | 0.00 | 33.80 | 45.19 | 0.00 | 43.20 | 44.94 | **8.00** | 29.71 | 36.61 | 0.00 |
| TransCoder | 37.71 | 39.35 | 0.00 | **25.99** | **34.39** | 0.00 | 19.91 | 30.09 | 0.00 | 31.75 | **48.37** | 0.00 | - | - | - | - | - | - |
| TransCoder-ST | **50.99** | **49.50** | **4.00** | 24.36 | 29.51 | 0.00 | 24.96 | 34.78 | **4.00** | 33.38 | 43.06 | 0.00 | - | - | - | - | - | - |

\* We omit the results of TransCoder(-ST) for Javascript since they do not support Javascript.

In conclusion, G-TransEval consists of parallel code pairs across five of the most popular programming languages: Python, C++, Java, C#, and JavaScript. By categorization, style normalization, and unit test case augmenting, G-TransEval provides a more comprehensive and accurate evaluation of code translation models, thereby enhancing the meaningfulness of performance comparisons between different approaches.

### B. Application

To showcase the usefulness of G-TransEval, we assess the performance of state-of-the-art code translation models using the proposed benchmark. Our objective is to uncover novel insights that have not been previously explored through comprehensive and in-depth comparisons of existing models. We train and evaluate all models according to the experimental setup in Section II-E, with CodeBERT and CodeT5 fine-tuned on the training set of XLCoST.

Table VI presents the translation performance of the models on the four types of translation. There are a total of 20 translation pairs spanning 5 popular languages. Due to space limitation, we only present the 6 groups involving the Java↔C++, Java↔Python and Java↔JavaScript translations. The full results are available in our Github repository [21].

*1) Effect of taxonomy:* We observe a clear trend across all translation pairs, where the translation performance decreases as the translation level increases. This is consistent with the findings in Section IV, which indicates that higher levels of translation pose greater difficulty for the models. Particularly, all models struggle with type-4 translations. This suggests that algorithm-level translation remains a significant challenge for current translation models.

Additionally, our analysis reveals that the performance of supervised and unsupervised approaches varies across different types of translations due to the varying knowledge required for each type. For example, TransCoder-ST, an unsupervised approach, achieves higher scores on type-1 translations, with a 7.9% higher accuracy (CA) compared to CodeT5 (supervised) in Java→C++ translation. This suggests that unsupervised methods can learn internal knowledge of programming languages through back-translation on large-scale code corpora, resulting in syntactically correct output. On the other hand, CodeT5 performs better on type-3 translations, with a 17.6% higher CA score compared to TransCoder-ST in Java→C++ translation. This implies that labeled data can provide external knowledge and improve model performance on API translation.

> **Finding 1:** G-TransEval with the taxonomy is effective in differentiating between various levels of translations. As the translation level increases, the task becomes more rigorous. It also demonstrates that unsupervised approaches exhibit better performance on lower levels, while supervised approaches demonstrate better performance on higher levels.

*2) Effect of programming languages:* Our benchmark study also indicates that the choice of programming language significantly impacts translation performance. Specifically, we categorize the five experimented languages into two groups: statically-typed (i.e., C++, Java, and C#) and dynamically-typed (i.e., Python and JavaScript), and examine the CA scores obtained by CodeT5 for all language pairs. We observe that type-1 translations within statically-typed languages yield higher accuracy (CA>90) than those across statically-typed and dynamically-typed languages (CA<90). This suggests that the more syntactically similar the languages are, the better performance for type-1 translations. Type-2 and type-3 translations across dynamically- to statically-typed languages exhibit more challenges. For example, Python (dynamic)→Java (static) and JavaScript (dynamic)→Java (static) translations receive the lowest scores among all translations. This is probably because dynamically-typed languages carry less information about variable types, which forces the model to infer variable types. We particularly notice that TransCoder-ST achieves significantly better performance than TransCoder on Java↔Python (static-dynamic) translations, implying the effectiveness of its self-training procedure in heterogeneous translations.

> **Finding 2:** The characteristics of languages have a strong impact on the performance under different translation types. Translations between syntactically dissimilar languages yield lower CA scores for type-1 translations. Similarly, translations from dynamically- to statically-typed languages are more challenging than other language pairs for type-2 and type-3 translations.

*3) Effect of stylization and diversification:* The results in Figure 2 show that supervised approaches achieve a significantly higher BLEU score than unsupervised approaches on the TransCoder-test benchmark. However, in G-TransEval, we observe the opposite result, where supervised approaches have inferior BLEU score compared to unsupervised counterparts in Java→C++ translation. This phenomenon could probably be attributed to the domain similarity: CodeBERT and CodeT5 are trained on the XLCoST corpus, which has the same source as TransCoder-test, while TransCoder and TransCoder-ST are trained using data collected from Github. Consequently, translations by CodeBERT and CodeT5 are more similar to the golden answers in terms of styles and naming conventions in the TransCoder-test benchmark. In contrast, G-TransEval mitigates this problem by normalizing the golden translations according to the Google code style, resulting in a fairer BLEU/CodeBLEU score.

Furthermore, we observe a significant decrease in the computational accuracy of type-2 and type-3 translations on the Java→C++ translation of G-TransEval, compared to the categorized TransCoder-test benchmark (Figure 2). As the model settings remain the same in both experiments, the result indicates that G-TransEval is inherently more challenging even without considering the type-4 translations. This could be attributed to the fact that G-TransEval is collected from multiple sources, with varying programming idioms and library usages.

> **Finding 3:** G-TransEval is more fairer and challenging on performance measures, possibly due to code style normalization and source diversification.

*4) Effect of training data size:* As code translation models are usually data-hungry, we also explore how different training

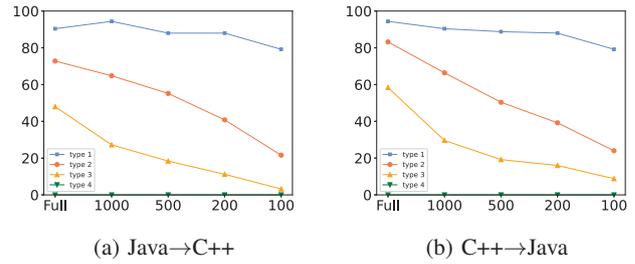

Fig. 4: Performance of CodeT5 under different training sizes on the four translation types, where the full dataset contains 8,389 training samples.

data sizes affect the performance across distinct translation types. We are particularly interested in how various models perform in the low-resource setting.

The results depicted in Figure 4 demonstrate that the size of the training samples has a detrimental impact on the translation performance under the four types. However, the degree to which each translation type is affected varies due to its different knowledge dependence. It is evident that reducing the training samples has a minimal impact on type-1 translations, as even with only 100 samples can still achieve comparable results to those trained on the full dataset.

The performance of type-2 and type-3 translations is significantly dependent on the training data size. For example, when only 1000 samples are used for training, the CA score witnesses a reduction of 11.0% and 43.3% for type-2 and type-3 translations, respectively. Furthermore, upon narrowing down to a mere 100 samples, the CA value plummets by 70.3% and 91.2% for the respective translation types.

Notably, we discern a 0% pass rate across all training data sizes for type-4 translations. Given the high complexity of this type, we posit that the existing training datasets prove inadequate for the construction of type-4 translation models. This underscores the greater dependence of type-4 translation on the quantity and quality of training datasets.

> **Finding 4:** The size of training data has various impacts on the translation performance, with type-1 translations more tolerant to small training data, followed by type-2, 3, and 4, which have increasing dependence on large-scale training data.

*5) Results of LLMs:* In addition to the state-of-the-art code translation models, we also evaluate large language models (LLMs) using our benchmark. Specifically, we experiment with two typical LLMs, including gpt-3.5-turbo [22] and StarCoderBase-15.5B [23]. We perform translation on gpt-3.5-turbo by leveraging the official API with default settings. A simple prompt template is used: "Translate the following [language A] function into [language B] function: [language A sample function]." As for StarCoderBase, we download the official checkpoint and perform the translation locally. For both LLMs, we generate one translation result for each test sample.

TABLE VII: Performance of LLMs on different translation types on the proposed benchmark (CB = CodeBLEU).

| Model | Java→Python | | | Python→Java | | | Java→C++ | | | C++→Java | | | Java→JavaScript | | | JavaScript→Java | | |
|---|---|---|---|---|---|---|---|---|---|---|---|---|---|---|---|---|---|---|
| | BLEU | CB | CA | BLEU | CB | CA | BLEU | CB | CA | BLEU | CB | CA | BLEU | CB | CA | BLEU | CB | CA |
| **Type 1** | | | | | | | | | | | | | | | | | | |
| gpt-3.5-turbo | 55.27 | 69.39 | **96.00** | 86.65 | 89.69 | **88.80** | 90.04 | 91.53 | **99.20** | 89.93 | 92.66 | **98.40** | **83.67** | **83.23** | **96.00** | 92.36 | **93.60** | **91.20** |
| StarCoderBase | **61.11** | **73.84** | 85.60 | **92.96** | **93.02** | 84.80 | **94.64** | **94.49** | 96.00 | **93.76** | **95.40** | 91.20 | 81.66 | 81.74 | 78.40 | 92.05 | 92.30 | 64.80 |
| **Type 2** | | | | | | | | | | | | | | | | | | |
| gpt-3.5-turbo | 54.78 | 67.08 | **93.60** | 74.72 | 72.20 | **84.80** | 83.10 | 83.72 | 87.20 | 84.06 | 82.68 | **88.80** | **85.86** | **84.31** | **91.20** | 76.41 | 73.59 | **80.80** |
| StarCoderBase | **61.48** | **72.61** | 83.20 | **87.13** | **84.62** | 79.20 | **91.96** | **91.03** | **93.60** | **90.35** | **88.75** | 79.20 | 84.51 | 83.32 | 88.00 | **88.54** | **85.53** | 63.20 |
| **Type 3** | | | | | | | | | | | | | | | | | | |
| gpt-3.5-turbo | 53.76 | 65.10 | **94.40** | 74.48 | 74.00 | **85.60** | 80.77 | 80.28 | **81.60** | 83.48 | 80.18 | **91.20** | **81.53** | **81.25** | **91.20** | 73.91 | 73.07 | **87.20** |
| StarCoderBase | **62.51** | **74.21** | 88.80 | **85.08** | **82.97** | 73.60 | **87.71** | **85.83** | 78.40 | **88.29** | **86.41** | 84.00 | 81.09 | 80.32 | 80.00 | **86.07** | **84.49** | 60.80 |
| **Type 4** | | | | | | | | | | | | | | | | | | |
| gpt-3.5-turbo | 27.59 | 40.42 | **72.00** | 37.60 | 47.81 | 64.00 | 35.57 | 43.08 | **68.00** | 44.73 | 55.14 | **68.00** | **61.67** | **61.99** | **76.00** | 35.62 | 47.78 | **88.00** |
| StarCoderBase | **41.14** | **47.91** | 44.00 | **54.09** | **59.55** | **72.00** | **40.61** | **44.95** | 48.00 | **60.73** | **64.25** | **68.00** | 49.29 | 47.52 | 56.00 | **55.72** | **62.60** | 64.00 |

The results are presented in Table VII. We observe that LLMs do not show a clear performance decrease under type-2 and type-3 translations as opposed to general PLMs. We believe that LLMs alleviate the knowledge gap of higher-level translations owing to the ultra-large code corpora used for pre-training. The large-scale data and model parameters enable LLMs to acquire sufficient syntactic and API knowledge that is required in type-2 and type-3 translations.

Comparing the two LLMs, gpt-3.5-turbo outperforms StarcoderBase in terms of CA on most translation pairs owing to the larger model size (gpt-3.5-turbo has approximately 11 times the parameters of StarCoderBase). But StarCoderBase also demonstrates strength over gpt-3.5-turbo in terms of BLEU and CodeBLEU, probably because these two metrics focus more on token-level similarities. Based on this observation, we believe that with the increase in model parameters and training data, larger models can acquire deeper knowledge of code, thus achieving competitive performance on higher levels of code translations.

> **Finding 5:** LLMs alleviate the knowledge gap of higher-level translations through the substantial number of parameters and training data, hence yielding competitive results in type-2 and type-3 translations.

## VI. DISCUSSION

Despite the demonstrated success of PLMs in code translation, our findings indicate that they have not yet reached their full potential. In this section, we delineate several future directions that could help advance the field.

*1) Building datasets for type-3 and type-4 translations:* It is observed that translations of type-3 and type-4 constitute a small fraction in prior benchmarks. Our experimental findings indicate that these translations are considerably more challenging than other types. As libraries are an essential component of software projects, type-3 and type-4 translations assume great significance and present significant obstacles to the development of a fully mature code migration system. Therefore, it is imperative that greater attention be devoted towards building datasets with type-3 and type-4 translations.

*2) Ensembling multiple pre-training approaches:* Our empirical findings indicate that different translation models specialize in acquiring different types of knowledge. Specifically, unsupervised methods that employ back-translation on a vast corpus may be more effective in capturing internal knowledge, while supervised models fine-tuned with parallel corpora may be more appropriate for acquiring external knowledge. However, neither approach is currently optimal for all types of translation. Consequently, combining multiple methods to leverage their respective strengths and mitigate their limitations holds great promise for enhancing overall translation performance.

*3) Cross-family program translation:* The findings in Section V-B2 reveal that the effectiveness of existing translation approaches is significantly impacted by the features of programming languages. Translation between languages of different families (e.g., static vs. dynamic, object-oriented vs. procedure, functional vs. imperative) is particularly challenging. As there are hundreds of programming languages, each having distinct features, it is crucial to investigate how these features affect the performance of translations and explore ways to enhance the translation quality for specific language pairs.

## VII. THREATS TO VALIDITY

*Internal validity.* One potential threat to the validity could be data leakage, that is, models may have already seen data in the benchmark during training, causing unfair comparisons. To mitigate this threat, we have conducted a deduplication procedure before the fine-tuning stage to make sure that there is no overlap between the training and test sets. Nevertheless, it is hard to filter out samples with subtle differences such as omitting braces. Moreover, models may have seen identical samples during the pre-training stage. In the future, we will provide a more thorough deduplication process for the benchmark data.

*External validity.* One external threat to the validity could be the model types we evaluated. Our empirical studies are carried out on four representative models, including CodeBERT, which is an encoder-only model; CodeT5, an encoder-decoder model; TransCoder, an unsupervised approach technique; and TransCoder-ST, the state-of-the-art unsupervised code translation model which utilizes unit tests to enhance the back-translation. Hence, the conclusions may not be generalized to LLMs. To mitigate this threat, we evaluate two LLMs on our benchmark, namely, StarCoderBase and gpt-3.5-turbo, and observe similar and explainable findings within the scope of our study. In the future, we will extend our study to the more extensive family of LLMs.

## VIII. Related Work

### A. Code Translation

Code translation plays a critical role in cross-language code migration. Recently, inspired by the great success of pre-trained models in NLP, a boom arises in pre-trained models on code migration. For example, CodeBERT [1], an encoder-only pre-trained model based on BERT [24], allows for learning NL-PL representations through an additional replaced token detection task [25]. Wang et al. proposed CodeT5 [3], which introduces pre-training tasks for code identifiers and comments based on T5 [13]. Since supervised approaches are costly in accumulating large parallel datasets [2], unsupervised approaches such as Transcoder [2] have been proposed and shown to be effective. More recently, TransCoder-ST [8] improves TransCoder by filtering out invalid program translations through unit tests in the back-translation stage.

With the advent of large-scale models, GPT [26] based models such as Codex [19], AlphaCode [27], PaLM-Coder [28], CodeGen [29], StarCoder [23], GPT-3.5 [22] and CodeGeeX [20] have also been applied to the field of code translation and have demonstrated exceptional performance.

### B. Code Translation Benchmarks

Besides our work, there have been other datasets and benchmarks in this field. Lu et al. [9] presented CodeXGLUE which contains parallel code from open-source projects. Puri et al. [30] built CodeNet, which covers 55 programming languages and includes code translation corpus collected from programming contest sites with extensive metadata. Zhu et al. [31] proposed a parallel code corpus called CoST, which covers seven programming languages and provides snippet-level alignment through code comment matching. They subsequently constructed a new dataset called XLCoST [14], which covers seven programming languages and natural language (English) and provides both snippet-level and program-level alignment. With the introduction of HumanEval [19], P3 [32], and MBPP [33], which only cover Python, various multilingual datasets have been created by extending Python code to other programming languages. For example, Zheng et al. [20] built HumanEval-X by manually writing solutions in C++, Java, JavaScript, and Go on HumanEval.

Our work differs from previous works in several ways. Firstly, we propose a 4-type code translation taxonomy and construct a new multilingual benchmark based on the taxonomy. Secondly, we conduct extensive experiments on the new benchmark using state-of-the-art models to obtain more empirical findings and a fine-grained comparison between these models. Our work offers a better understanding of the strengths and limitations of existing code translation models and provides new insights for improving model development in the future.

### C. Complexity Categorization for Source Code

To the best of our knowledge, we are the first to develop a taxonomy of code translation tasks. Besides our work, we also notice similar code taxonomy works in the software engineering area. For example, Roy et al. [34] categorized code clones into four types. Their taxonomy has been widely accepted and utilized to evaluate the strength and weaknesses of different techniques used in code clone detection area [35].

Compared to the code clone taxonomy approach, which focuses on detecting transformations within the same programming language, our approach takes into account differences in language features, such as syntactic structures and API correspondences, across different programming languages.

Similarly, Hendrycks et al. [18] collected a code synthesis dataset called APPS from open-access sites including Codewars, AtCoder, Kattis, and Codeforces. They categorized the samples in this dataset into 3 levels: introductory level, interview level, and competition level, according to their difficulties for solving. Our taxonomy differs significantly from theirs as code translation relies on different knowledge from those for code synthesis.

## IX. Conclusion

In this paper, we evaluate the fine-grained ability of code translation models in six dimensionalities and summarize the translation features across different score ranges. Based on the empirical findings, we develop a taxonomy of code translation tasks with four types that increase in complexity and knowledge dependence. Based on the taxonomy and empirical findings, we propose a new benchmark called G-TransEval by carefully curating type-3 and type-4 translations and unit test cases. Experiments on the new benchmark show that G-TransEval provides a more fine-grained evaluation of model capabilities, leading to several insights and findings that could help improve the development and evaluation of code translation systems. In the future, we plan to expand the proposed benchmark to include more high-quality data samples. Moreover, we will collect a categorized training set for developing more advanced code translation models.

We have released the whole suite of our benchmark and the full experimental results online [21].

## Acknowledgment

This research is supported by the National Natural Science Foundation of China (Grant No. 62232003, 62102244, 62032004) and CCF-Tencent Open Research Fund (RAGR20220129).